\documentclass[12pt]{iopart}
\usepackage{graphicx}
\usepackage{amssymb}

\newcommand{\btau}{\mbox{\boldmath{$\tau$}}}

\newcommand{\bU}{\mbox{\boldmath{$U$}}}

\newcommand{\bH}{\mbox{\boldmath{$H$}}}

\newcommand{\bb}{\mbox{\boldmath{$b$}}}
\newcommand{\bbr}{\mbox{\boldmath{$r$}}}

\newcommand{\bone}{\mbox{\boldmath{$1$}}}

\newcommand{\bv}{\mbox{\boldmath{$v$}}}

\newcommand{\bx}{\mbox{\boldmath{$x$}}}

\newcommand{\bQ}{\mbox{\boldmath{$Q$}}}
\newcommand{\bZ}{\mbox{\boldmath{$Z$}}}
\newcommand{\bL}{\mbox{\boldmath{$L$}}}

\newcommand{\be}{\mbox{\boldmath{$e$}}}
\newcommand{\bR}{\mbox{\boldmath{$R$}}}
\newcommand{\bO}{\mbox{\boldmath{$O$}}}
\newcommand{\by}{\mbox{\boldmath{$y$}}}

\newcommand{\bu}{\mbox{\boldmath{$u$}}}
\newcommand{\bI}{\mbox{\boldmath{$I$}}}

\newcommand{\bsH}{\mbox{\scriptsize \boldmath{$H$}}}
\newcommand{\bsQ}{\mbox{\scriptsize \boldmath{$Q$}}}
\newcommand{\bsb}{\mbox{\scriptsize \boldmath{$b$}}}
\newcommand{\bstau}{\mbox{\scriptsize \boldmath{$\tau$}}}

\newcommand{\boeta}{\mbox{\boldmath{$\eta$}}}
\newcommand{\mC}{\mathbb{C}}
\newcommand{\mR}{\mathbb{R}}
\newcommand{\mN}{\mathbb{N}}

\begin{document}
\title{Statistical mechanical analysis 
of the linear vector channel in digital communication }

\author{Koujin Takeda\dag, Atsushi Hatabu\dag\ddag\ and Yoshiyuki
Kabashima\dag}
\address{\dag \ Department of Computational Intelligence and Systems Science, Tokyo
Institute of Technology, Yokohama 226-8502, Japan \\
\ddag \ System IP Core Research Laboratories, NEC Corporation,
1753 Shimonumabe, Nakahara, Kawasaki 211-8666, Japan}

\begin{abstract}
A statistical mechanical framework to analyze linear vector 
channel models in digital wireless communication 
is proposed for a large system. 
The framework is a generalization of that proposed for 
code-division multiple-access
systems in \textit{Europhys. Lett.,}~\textbf{76}
 (2006) {1193} and enables the analysis of the 
system in which the elements of the channel transfer matrix
are statistically correlated with each other. The significance of the proposed scheme 
is demonstrated by assessing the performance of an existing model 
of multi-input multi-output communication systems.
\end{abstract}


%%%%%%%%%%%%%%%%%%%%%%%%%%%%%%%%%
\section{Introduction}
In recent years, the number of objects to which statistical mechanical analysis
can be applied has increased rapidly.
The digital wireless communication system is
one such example, and many intriguing studies in this field
have revealed a strong relationship between telecommunication 
systems and statistical mechanics \cite{GuoVerdu2005,Nishimori2001}. 

The linear vector channel is one of the basic categories of
wireless communication system.  
Code division multiple access (CDMA) 
\cite{Verdu1998}, which is employed in third-generation 
cellar phone systems and wireless LANs, is a type of linear vector channel. 
In the general CDMA scenario, many users transmit 
discrete symbols that are modulated by random signature sequences
using a single channel, and mixtures of user signals and noises 
are received at the other end of the channel. 
This indicates that the problem of simultaneously demodulating 
user signals from received signals can be mapped to a virtual 
spin system governed by random interactions. 
This problem has been successfully solved by techniques developed 
in statistical mechanics for disordered systems, and in particular by
the replica method \cite{Tanaka2001,Tanaka2002,Wenetal2005,Wen2006,Guo2006}. 

The multi-input multi-output (MIMO) system is another well-known example
of a linear vector channel to which the statistical mechanical
approach is applicable \cite{Moustakas2003,Muller2003,TakeuchiTanakaYano2007}.
A MIMO system is composed of many transmit and receive antennas. 
In a general scenario, multiple input signals transmitted 
from transmit antennas are received at the receive antennas, 
being linearly transformed to multiple output signals 
by a channel transfer matrix. In several preceding studies, 
channel transfer matrices are regarded as deterministic. 
However, the elements of such matrices 
vary with time in actual cases, which implies that handling 
the matrices as random is more realistic. In the simplest model, 
each element of the matrix could be regarded as an independent Gaussian 
variable with zero mean. Unfortunately, modeling of this type is 
inadequate for describing realistic MIMO systems 
in which correlations among the matrix elements 
are, in general, not negligible due to 
spatial proximity among transmit or receive antennas. 
For continuous inputs that are modeled as Gaussian variables, 
simple expressions of 
performance evaluation can still be obtained by using knowledge of 
random matrix theory \cite{TulinoVerdu2004}. 
However, such expressions cannot be applied directly to 
discrete inputs, which are usually used in digital communication. 
Therefore, developing a framework to analyze MIMO systems, 
and, more generally, linear vector channels of discrete inputs
are demanded. 

The purpose of the present article is to meet such a demand. 
Recently, the authors proposed a scheme to analyze 
CDMA systems under the assumption that a cross-correlation 
matrix of signature sequences can be regarded as a sample generated from 
a certain type of random matrix ensemble, which is 
characterized by an eigenvalue spectrum \cite{TakedaUdaKabashima2006}.
We herein generalize this scheme so as to be 
applicable to a wider class of linear vector channels.

The present paper is organized as follows. In Section 2, the linear vector channel models investigated herein are introduced. Section 3, in which a framework to analyze a given system is developed based on the replica method, is the main part of this article. The assumption of uniformity of transmitted signals generally guarantees that the average of the replicated Boltzmann weight depends on replicated vectors 
only through overlaps among the replicated and original vectors. This makes it possible to evaluate typical properties of the target system using a single function, which is referred to as $G(x)$ \cite{MarinariParisiRitort1994,ParisiPotters1995}. 
In general, the analysis of typical property requires assessment of quenched averages, which implies that $G(x)$ should be evaluated as a quenched average utilizing 
the replica method. However, we will show that the assessment of this function can generally be reduced to the calculation of an annealed average due to a distinctive property underlying the evaluation of the average eigenvalue spectrum of the cross-correlation matrix for a given channel transfer matrix ensemble using the replica method if the eigenvalue spectrum of the ensemble is self-averaging. In Section 4, the significance of the framework is demonstrated by application to one of the typical MIMO models called the Kronecker model. Finally, Section 5 presents a summary of the present study.

%%%%%%%%%%%%%%%%%%%%%%%%%%%%%%%%%
\section{Model definition}
A linear vector channel is defined as a system 
in which an input vector composed of $K$ components, $\bb=(b_k)$
$(k=1,2,\ldots,K)$ (boldface denotes vector or matrix), 
is linearly transformed by an $L \times K$ 
channel transfer matrix $\bH$ and is additively degraded by noise. 
For generality and simplicity, we assume that $\bH$ and 
$\bb$ are defined over the complex number field, and 
the channel noise is given as circularly
symmetric complex additive white Gaussian noise,
the variance of which is $N_0$. 
We denote ${\rm Re}(u)$ and ${\rm Im}(u)$ as real and complex parts 
of a complex number $u$, respectively.
$|u|=\sqrt{{\rm Re}(u)^2+{\rm Im}(u)^2}$ denotes 
the absolute value of $u$. 
Under these assumptions and notations, the output vector 
\begin{eqnarray}
\bbr= \bH \bb^0 +\sqrt{N_0} \boeta, 
\label{received_signal}
\end{eqnarray}
is received by a receiver, where
$\bb^0$ denotes the input vector that is actually transmitted,
and the components of the noise vector $\boeta$, $\eta_l$ ($l=1,2,
\ldots,L$), independently obey circularly symmetric 
complex normal distributions $P(\eta) = \pi^{-1}
\exp \left [-|\eta|^2 \right ] =
\pi^{-1}\exp \left [-({\rm Re}(\eta)^2+{\rm Im}(\eta)^2)
\right ] $. 

In the performance analysis shown below, $\bH$ is regarded as a sample from a certain 
random matrix ensemble, typical samples of which are {\em dense}. 
Namely, we assume that most elements of typical $\bH$ do not vanish. 
Since an elegant framework has been already established 
for Gaussian inputs \cite{TulinoVerdu2004}, 
we focus on cases of discrete inputs in which input symbols 
are expressed as $b_k \in \{e^{2 \pi i s/S}| s=0,1,2, \ldots, S-1\}
\equiv {\cal A}_S$, where $i=\sqrt{-1}$. 
This expression corresponds to standard digital communication schemes 
of binary phase shift-keying (BPSK) and quadratic phase 
shift-keying (QPSK) for $S=2$ and $S=4$, respectively. 
We further assume that $\bb$ is encoded so as to 
be uniformly generated as $P(\bb)=1/S^{K}$ for optimizing 
communication performance. 

After receiving $\bbr$ the remaining task for the receiver is 
to infer the original vector $\bb^0$. 
The optimal inference scheme to minimize the component-wise
probability of incorrect estimation, which is referred to as 
$P_b$, is constructed from the posterior distribution 
\begin{eqnarray}
P(\bb|\bbr)=Z^{-1}
\exp \left [
-\frac{1}{N_r}
|\bbr - \bH \bb|^2
\right ], 
\label{posterior}
\end{eqnarray}
as $\hat{b}_k=\mathop{\rm argmax}_{b_k}\left \{\sum_{\bsb 
\backslash b_k  } 
P(\bb|\bbr) \right \}$
in the hope that a model parameter of 
noise variance $N_r$ is in agreement with the correct value $N_0$.
Here, 
\begin{equation}
  Z \equiv \sum_{\bsb \in {\cal A}_S^K}\exp \left [
-\frac{1}{N_r}|\bbr- \bH \bb|^2
 \right ]
\label{partition_function}
\end{equation}
serves as the partition function, 
and $\hat{b}_k$ denotes the estimate of $b_k^0$, 
where ${\cal A}_S^K$ denotes the $k$-th extension of 
symbol set ${\cal A}_S$. 

%%%%%%%%%%%%%%%%%%%%%%%%%%%%%%%%%
\section{Analytical scheme}
\subsection{Gauge transformation and Haar measure}
Since the posterior distribution (\ref{posterior}) depends on 
predetermined random variables $\bH$, $\bb^0$, and $\boeta$, we 
resort to the replica method for assessing typical 
property of the linear vector channel. 
Thus, we first substitute equation (\ref{received_signal})
into equation (\ref{partition_function}) and 
perform gauge transformation $(b_k^0)^*b_k^a \to \tau_k^a$ 
$(k=1,2,\ldots,K; a=1,2,\ldots,n)$, where $*$ denotes 
the complex conjugate. This yields an expression of 
the replicate partition function (\ref{partition_function}) 
for $n=1,2,\ldots$ as
\begin{eqnarray}
Z^n =\sum_{\bstau^1,\bstau^2,\ldots,\bstau^n\in {\cal A}_S^K}
\exp \left [ -\frac{1}{N_r } \sum_{a=1}^n 
|\bH {\rm diag}(\bb^0)(\bone -\btau^a) +\sqrt{N_0}\boeta|^2 
\right ], 
\label{gauged_form}
\end{eqnarray}
where $\bone$ is a $K$-dimensional vector, all elements of which 
are unity and $\btau^a=(\tau_k^a)$ $(k=1,2,\ldots,K; a=1,2,\ldots,n)$. 
The diagonal matrix ${\rm diag}(\bb^0)$ is 
defined as ${\rm diag}(\bb^0)\equiv (\delta_{kj} b_k^0)$. 
Next, we average this expression 
with respect to $\bb^0$ over the correct prior $P(\bb^0)=1/S^K$, 
fixing gauged vectors $\{\btau^a\}=\{\btau^1,\btau^2,\ldots, \btau^n\}$.  
Here, we consider a property whereby
vectors ${\rm diag}(\bb^0)(\bone -\btau^a)$ 
$(a=1,2,\ldots,n)$ are sampled isotropically in 
$K$-dimensional vector space under constraints of 
relative configuration 
\begin{eqnarray}
(\bone-\btau^a) \cdot (\bone-\btau^b)
= (\bb^0-\bb^a) \cdot (\bb^0-\bb^b) 
= K(1-m_a^*-m_b+q_{ab}), \label{constraints1} 
\end{eqnarray}
for $a\ne b(=1,2,\ldots,n)$, and
\begin{eqnarray}
(\bone-\btau^a) \cdot (\bone-\btau^a)
= (\bb^0-\bb^a) \cdot (\bb^0-\bb^a) 
= 2K(1-m_a), 
\label{constraints2} 
\end{eqnarray}
for $a=1,2,\ldots,n$, 
where $\cdot$ is defined as 
$\bx \cdot \by=\sum_{k=1}^K x_k^* y_k$
for complex vectors $\bx=(x_k)$ and $\by=(y_k)$ 
$(k=1,2,\ldots,K)$. 
$m_a=(1/K)\bone \cdot \btau^a=(1/K)\bb^0 \cdot \bb^a$ 
and $q_{ab}=(1/K)\btau^a \cdot \btau^b=(1/K)\bb^a \cdot \bb^b
=q_{ba}^*$
($a,b = 1,2,\ldots,n$). 
This implies that for any fixed set of $\{\btau^a\}$, 
the average of the replicated Boltzmann weight
$\exp \left [ -1/( N_r) \sum_{a=1}^n 
|\bH {\rm diag}(\bb^0)(\bone -\btau^a) +\sqrt{N_0}\boeta|^2 
\right ]$ with respect to $\bb^0$ is assessed as
\begin{eqnarray}
&&\frac{1}{S^K}\sum_{\bsb^0 \in {\cal A}_S^K}
\exp \left [ -\frac{1}{ N_r} \sum_{a=1}^n 
|\bH {\rm diag}(\bb^0)(\bone -\btau^a) +\sqrt{N_0} \boeta|^2 
\right ]\cr
&&\simeq \int {\cal D} \bU
\exp \left [ -\frac{1}{ N_r} \sum_{a=1}^n 
|\bH \bU (\bone -\btau^a) +\sqrt{N_0} \boeta|^2 \right ], 
\label{replacement}
\end{eqnarray}
where ${\cal D}\bU$ denotes the Haar measure
of unitary matrix $\bU$, which is normalized
as $\int {\cal D}\bU=1$. 

Equation (\ref{replacement}) is justified by the 
following arguments. For a fixed typical pair 
of $\{\btau^a\}$ and $\bH$, which is dense,
components of $\bH {\rm diag}(\bb^0) (\bone -\btau^a)$, 
denoted as $t_l^a=\sum_{k,j}H_{l k} b_k^0 \delta_{kj}(1-\tau_j^a)$
$=\sum_{k}H_{lk} b_k^0 (1-\tau_k^a)$, 
are composed of many independent random variables
and, therefore, can be dealt with as complex Gaussian 
variables as a consequence of the central limit theorem 
when $\bb^0$ is sampled from the uniform distribution $P(\bb^0)=1/{\cal S}^K$.
This means that statistical properties of $t_l^a$ are
fully characterized by only the first and second moments, 
which are determined by those of matrix components of ${\rm diag}(\bb^0)$. 
More precisely, the relevant moments are evaluated as
$[t_l^a]_{\bsb^0}=0$
and $[(t_l^a)^* t_m^b]_{\bsb^0}
=\sum_{k,n}
H_{l k}^*H_{m n} [(b_k^0)^*b_n^0]_{\bsb^0} (1-\tau_k^a)^*
(1-\tau_n^b)$
$=\sum_{k,n}
H_{l k}^*H_{m n} \delta_{kn} (1-\tau_k^a)^*
(1-\tau_n^b)$
$=\sum_{k}H_{l k}^*H_{m k} (1-\tau_k^a)^*(1-\tau_k^b)$
$\simeq (\sum_{k}H_{l k}^*H_{m k})\times 
K^{-1}(\bone-\btau^a)\cdot (\bone-\btau^b)$
for $l,m=1,2,\ldots,L$ and $a,b=1,2,\ldots,n$, 
where $[\cdots ]_{\bsb^0}$ represents average with respect to 
$P(\bb^0)=1/{\cal S}^K$. The final replacement for the second moments
$[(t_l^a)^* t_m^b]_{\bsb^0}$ is allowed as $\bH$ and 
$\{\btau^a\}$ are statistically uncorrelated a priori. 
Here, it is noteworthy that the identical moments are reproduced by 
substituting unitary matrix $\bU$ for ${\rm diag}(\bb^0)$
in conjunction with replacement of the Haar measure ${\cal D} \bU$
with $P(\bb^0)=1/{\cal S}^K$. 
This validates equation (\ref{replacement}), which will be supported 
by numerical experiments shown later herein as well. 

\subsection{$G$-functions and free energy}
Next, averaging with respect to $\boeta$, we obtain 
\begin{eqnarray}
&& \int_{\mC^{L}} \frac{d \boeta }{\pi^L} 
\exp \left [-|\boeta|^2 \right ]
\times \exp \left [ -\frac{1}{ N_r} \sum_{a=1}^n 
|\bH \bU (\bone-\btau^a) +\sqrt{N_0} \boeta|^2 \right ]  \cr
&&= \exp \left [ {\rm Tr}
{\bR} \bL(n)
\right ], 
\label{Ln}
\end{eqnarray}
where 
$d \boeta=\prod_{l=1}^L d{\rm Re}(\eta_l)\prod_{l=1}^L d{\rm Im}(\eta_l)$, 
$\bR=\bU^\dagger(\bH^\dagger \bH )\bU$ and 
\begin{eqnarray}
\bL(n)=&-&\frac{1}{N_r}
\sum_{a=1}^n (\bone-\btau^a)(\bone-\btau^a)^\dagger \cr
&+&\frac{ N_0}{N_r(N_r+n N_0)}
\left (\sum_{a=1}^n (\bone-\btau^a) \right )
\left (\sum_{b=1}^n (\bone-\btau^b) \right )^\dagger, 
\label{operator}
\end{eqnarray}
%where $\dagger$ denotes the Hermitian conjugate. 
where $\int_{{\cal X}}$ denotes integration over a certain support set 
${\cal X}$. 
Next, integrating equation (\ref{Ln})
over the Haar measure $ {\cal D}\bU$
in conjunction with taking average with respect to 
$\bH$ yields an expression of the averaged Boltzmann factor: 
\begin{eqnarray}
&&\overline{\exp \left [ -\frac{1}{N_r } \sum_{a=1}^n 
|\bH {\rm diag}(\bb^0)(\bone -\btau^a) +\sqrt{N_0} \boeta|^2 
\right ]} \cr
&& \simeq
\left [\int {\cal D} \bU \exp \left [ 
{\rm Tr}\bR \bL(n) \right ] \right ]_{\bsH} \cr
&& \simeq 
\left [\exp \left [ K {\rm Tr} G_{\bsH}\left (\frac{\bL(n)}{K}
\right ) \right ] \right ]_{\bsH} 
\simeq 
\exp \left [ K {\rm Tr} \left [G_{\bsH}\left (\frac{\bL(n)}{K}
\right )\right ]_{\bsH} \right ]  
\label{anneal_quench_G} \\
&& \equiv \exp \left [ K {\rm Tr} G\left (\frac{\bL(n)}{K}
\right ) \right ], 
\label{G_function}
\end{eqnarray}
for large $K$, where $\overline{\cdots}$ and $\left [ \cdots \right ]_{\bsH}$
denote the averages with respect to $\bb^0$, $\boeta$, and $\bH$
and to only $\bH$, respectively. 
Transformation of equation (\ref{anneal_quench_G}) is valid
if the eigenvalue spectrum of $\bH^\dagger \bH$ is self-averaging, {\em i.e.}, if the discrepancy between the eigenvalue spectrum of typical samples
of $\bH^\dagger \bH$ and its average vanishes as $K,L \to \infty$
with keeping the load $\beta=K/L$ finite, 
as assumed hereinafter. 
A practical method to evaluate 
functions $G_{\bsH}(x)$ and $G(x)$ and the validation 
of equation (\ref{anneal_quench_G}) are discussed later herein. 

Intrinsic permutation symmetry among replicas naturally 
leads to the replica symmetric (RS) ansatz.
This implies that configurations characterized 
by $\bone \cdot \btau^a=\bb^0 \cdot \bb^a=K m$ $(a=1,2,\ldots,n)$ 
and $\btau^a\cdot \btau^b=\bb^a \cdot \bb^b =K q$ $(a \ne b)$
provide the most dominant contribution to the evaluation
of $\overline{Z^n}$. 
Note that the RS ansatz requires obvious symmetry 
$\btau^a \cdot \btau^b=\btau^b \cdot \btau^a$, 
which enforces $q_{ab}=q_{ba}=q$ to be {\em real} at the saddle point
for $\forall{a,b=1,2,\ldots,n}$. 
Under this ansatz, the $K \times K$ matrix $\bL (n)$ 
has three types of eigenvalues: $\lambda_1=-K(N_r+n N_0)^{-1}
(1-q+n(1-2m+q))$, 
$\lambda_2=-KN_r^{-1}(1-q)$ and $\lambda_3=0$, 
the numbers of degeneracy of which are 
$1$, $n-1$, and $K-n$, respectively. 
This indicates that equation (\ref{G_function}) is evaluated as
\begin{eqnarray}
\exp \left [ 
K \left (
G \left (-\frac{1-q+n(1-2m+q)}{N_r+n N_0} \right )
+(n-1) G\left (-\frac{1-q}{N_r} \right ) \right ) \right ]. 
\label{eigenvalues}
\end{eqnarray}
In addition, the RS ansatz offers
the number of microscopic configurations that satisfy 
constraints (\ref{constraints1}) and (\ref{constraints2}) as 
\begin{eqnarray}
\mathop{\rm Tr}_{\{\bstau^a\}}
\prod_{a=1}^n \delta(\bone \cdot \btau^a-Km)
\prod_{a>b} \delta(\btau^a \cdot \btau^b-Kq) \simeq 
\exp \left [K S_n(q,m) \right ], 
\label{entropic}
\end{eqnarray}
where 
\begin{eqnarray}
&&S_n(q,m)=
\mathop{\rm Extr}_{\hat{q},\hat{m}}\left \{
\ln \left [
\int_{\mC} D\zeta \left (
\sum_{\tau \in {\cal A}_S}
\exp 
\left [ {\rm Re}\left ((\sqrt{\hat{q}} \zeta^* + \hat{m})
\tau 
\right ) \right ] \right )^n \right ] \right . \cr
&& \left . 
\phantom{S_n(q,m)={\rm Extr}_{\hat{q},\hat{m}}\{aaaaaaa}
- n\hat{m}m-\frac{n}{2}\hat{q}-\frac{n(n-1)}{2}\hat{q}q \right \}, 
\end{eqnarray}
where $\mathop{\rm Extr}_{u} \{ \cdots  \}$ indicates 
the extremization of $\{ \cdots \}$ with respect to $u$
and $D\zeta=(d {\rm Re}(\zeta) d {\rm Im}(\zeta)/2 \pi)
\exp \left [-|\zeta|^2/2 \right ]$ denotes the complex Gaussian measure,
the variance of which is normalized to unity in each direction of 
the real and complex axes. 
Analytically continuing equations (\ref{eigenvalues})
and (\ref{entropic}) from $n \in \mN$ to $n \in {\bf \mR}$ yields
an expression for assessing the configurational average of free energy 
as \begin{eqnarray}
&&\frac{1}{K} \overline{\ln Z} =\lim_{n \to 0}
\frac{1}{nK} \ln \overline{Z^n} \cr 
&&=\mathop{\rm Extr}_{m,q,\hat{m},\hat{q}}
\left \{
G\left (-\frac{1-q}{N_r} \right )
+\left (-\frac{1-2m+q}{N_r}+\frac{N_0(1-q)}{N_r^2}
\right )
G^\prime \left (-\frac{1-q}{N_r} \right ) \right . \cr
&& \left . -\hat{m}m-\frac{\hat{q}(1-q)}{2} 
+ \int_{\mC} D\zeta \ln \left (
\sum_{\tau \in {\cal A}_S}
\exp \left [ {\rm Re}\left ((\sqrt{\hat{q}} \zeta^* + 
\hat{m})\tau 
\right ) \right ]
\right )
\right \}. 
\label{free_energy}
\end{eqnarray}
The saddle-point solution yields the bit error rate for
the demodulation strategy 
$\hat{b}_k=\mathop{\rm argmax}_{b_k}\left \{\sum_{\bsb 
\backslash b_k  } 
P(\bb|\bbr) \right \}$ as
$P_b=\int_{\mC} D\zeta \Theta_{\rm error}(\zeta;\hat{q},\hat{m})$, 
where $\Theta_{\rm error}(\zeta;\hat{q},\hat{m})$ vanishes
if $\exp \left [ {\rm Re}\left ((\sqrt{\hat{q}} \zeta^* + 
\hat{m})\tau 
\right ) \right ]$ is maximized by $\tau=1$ 
among $\tau \in {\cal A}_S$, and unity, otherwise. 

Note that both the RS assessment presented 
here and that of the replica symmetry breaking (RSB) ansatz are 
generally possible following the current framework. 
For instance, as shown in Appendix A, 
one can evaluate the free energy under the ansatz 
of one step replica symmetry breaking (1RSB) by dividing 
$n$ replicated vectors $\{ \btau^a\}$ into $n/x$ groups 
of identical size $x$ and assuming that 
the correct saddle point is characterized by 
the following relative configuration: $\btau^a \cdot \btau^b=K, K(q+\Delta)$, and 
$Kq$ for $a=b$, for the case in which $a \ne b$ 
but $a$ and $b$ belong to an identical 
group and otherwise, respectively. 
In the limit $n \to 0$, the 1RSB saddle-point equation is obtained as 
\begin{eqnarray}
\begin{array}{l}
\hat{\Delta}=\frac{2}{x N_r}  (G^\prime_0-G^\prime_1), \quad
\hat{q}=\frac{2N_0}{N_r^2}G^\prime_1-\frac{2 A}{N_r} 
G^{\prime \prime}_1, 
\quad \hat{m}= \frac{2}{N_r} G_1^\prime,  \cr
\Delta=\int_{\mC} D\zeta 
\left (\frac{\int_{\mC} D\eta \Xi^x |\left \langle \tau \right \rangle_1|^2}
{\int_{\mC} D\eta \Xi^x} 
-\left |\frac{\int_{\mC} D\eta \Xi^x \left \langle \tau \right \rangle_1}
{\int_{\mC} D\eta \Xi^x} \right |^2 \right ), \cr
q= \int_{\mC} D\zeta \left 
|\frac{\int_{\mC} D\eta \Xi^x \left \langle \tau \right \rangle_1}
{\int_{\mC} D\eta \Xi^x} \right |^2, \cr
m= \int_{\mC} D\zeta \frac{\int_{\mC} D\eta \Xi^x 
{\rm Re}(\left \langle \tau \right \rangle_1)}
{\int_{\mC} D\eta \Xi^x}. 
\end{array}
\label{1RSB_SP}
\end{eqnarray}
Here, 
$G_1^\prime=G^\prime(-(1-q+(x-1)\Delta)/N_r)$, 
$G_1^{\prime \prime}=G^{\prime \prime}(-(1-q+(x-1)\Delta)/N_r)$, 
$G_0^\prime=G^\prime(-(1-q-\Delta)/N_r)$, 
$A=-(1-2m+q)/N_r+(N_0/N_r^2)(1-q+(x-1)\Delta)$, 
$\Xi=\sum_{\tau \in {\cal A}_S}
\exp \left [ {\rm Re}\left (
(\sqrt{\hat{\Delta}} \eta^*
+\sqrt{\hat{q}} \zeta^* + \hat{m}
)
\tau \right ) \right ]$, 
and 
$\left \langle \tau \right \rangle_{1}
=\Xi^{-1} 
\sum_{\tau \in {\cal A}_S} \tau 
\exp \left [ {\rm Re}\left (
(\sqrt{\hat{\Delta}} \eta^*+\sqrt{\hat{q}} \zeta^* + 
\hat{m})\tau 
\right ) \right ]$. 

For $x \to 1$, a useful relation 
\begin{eqnarray}
\frac{\int_{\mC} D\eta \Xi \left \langle \tau \right \rangle_1}
{\int_{\mC} D\eta \Xi }=
\frac{\sum_{\tau \in {\cal A}_S}\tau 
\exp \left [ {\rm Re}\left ((\sqrt{\hat{q}} \zeta^* + 
\hat{m})\tau 
\right ) \right ]}
{\sum_{\tau \in {\cal A}_S}
\exp \left [ {\rm Re}\left ((\sqrt{\hat{q}} \zeta^* + 
\hat{m})\tau 
\right ) \right ]} \equiv \left \langle \tau \right \rangle_0, 
\label{1RSB_RS}
\end{eqnarray}
implies that the saddle-point condition 
of equation (\ref{1RSB_SP}) is governed by only four out of six
variables, {\em i.e.}, $\hat{q},\hat{m}, q$ and $m$. 
Actually, the condition for determining the four variables
in this case corresponds to the saddle-point equation of RS free energy 
(\ref{free_energy}), which implies that $x \to 1$ RSB analysis 
is generally reduced to that of RS. Nevertheless, 
a nontrivial result can still be obtained by investigating the behaviors of $\hat{\Delta}$ and $\Delta$, 
which are subserviently determined from equation (\ref{1RSB_SP}). 
This equation guarantees that $\hat{\Delta}=\Delta=0$ 
always satisfies the saddle-point condition. 
However, for $x \to 1$ stability analysis indicates that a nontrivial 
solution of $\hat{\Delta} >0$ and $\Delta>0$ emerges if
\begin{eqnarray}
\frac{2}{N_r^2}G^{\prime \prime}
\left (-\frac{1-q}{N_r} \right )
\int_{\mC} D \zeta 
\left (1-|\left \langle \tau \right \rangle_0|^2 \right )^2 > 1, 
\label{AT}
\end{eqnarray}
which is in accordance with the de Almeida-Thouless (AT) 
condition signaling the local instability of the RS solution \cite{AT1978}.

\subsection{Equivalence between quenched and annealed averages
in the assessment of $G(x)$ }
$G_{\bsH}(x)$ can be assessed by several formulae \cite{OpperWinther2001L,OpperWinther2001,CherrierDeanLefevre2003}. 
One formula uses the Stieltjes (or Cauchy) transformation of 
$\rho_{\bsH}(\lambda)=(1/K) \sum_{k=1}^K \delta (\lambda-\lambda_k)$, which 
is the eigenvalue spectrum of cross-correlation matrix $\bH^\dagger \bH$, 
\begin{eqnarray}
x=\int \frac{d \lambda \rho_{\bsH}(\lambda) }{\Lambda(x)-\lambda}
\label{cauchy}
\end{eqnarray}
where $\lambda_1,\lambda_2,\ldots,\lambda_K$ are
eigenvalues of $\bH^\dagger \bH$ and are guaranteed to be real
because $\bH^\dagger \bH$ is Hermitian. 
For given $x$, this relationship determines $\Lambda(x)$ implicitly, 
which is termed the Stieltjes inversion formula. 
Using $\Lambda (x)$, $G_{\bsH}(x)$ is assessed as
\begin{eqnarray}
G_{\bsH}(x)=\int_0^x dt \left (\Lambda (t)- t^{-1} \right ), 
\label{R_transform}
\end{eqnarray}
which is equivalent to the $R$-transformation known in free probability 
theory \cite{VoiculescuDykemaNica1992,MullerGuoMoustakas2007}. 

Here, we describe a rather primitive approach. 
For this objective, we substitute $x \be \be^\dagger$ 
for $\bL(n)$ in equation (\ref{Ln}), where $\be$ is a certain complex vector,
the length of which is fixed as $|\be|^2=K$. 
Note that the eigenvalues of this operator are $K x$ and zero, 
the degeneracies of which are $1$ and $K-1$, respectively. 
Integrating over ${\cal D}\bU$ yields the following expression: 
\begin{eqnarray}
&&\exp \left [ KG_{\bsH}(x) \right ]
= \exp \left [ K {\rm Tr}  G_{\bsH}\left (\frac{x\be \be^\dagger}{K} \right ) 
\right ]  =\int {\cal D}\bU \exp \left [{\rm Tr}\bR (x \be \be^\dagger) 
\right ] \cr
&& =\frac{\int_{\mC^{K}} d \bu \delta(|\bu|^2-Kx) \exp \left [\bu^\dagger
(\bH^\dagger\bH) \bu \right ]}
{\int_{\mC^{K}} d \bu \delta(|\bu|^2-Kx)}, 
\label{spherical}
\end{eqnarray}
where we set $\bu=\sqrt{x} \bU \be$. 
Inserting $\delta(|\bu|^2-Kx)=\int_{-i\infty}^{+i\infty}
d \Lambda \exp \left [-\Lambda(|\bu|^2-Kx) \right ]/(2 \pi i)$ 
and employing the saddle-point method with respect to the integration of 
$\Lambda$ yields the following expression: 
\begin{eqnarray}
G_{\bsH}(x)&=&\mathop{\rm Extr}_{\Lambda}\left \{
-\frac{1}{K} \ln \det |\Lambda-\bH^\dagger\bH| +\Lambda x \right \}
-\ln x -1, \cr
&=& \mathop{\rm Extr}_{\Lambda}\left \{
-\frac{1}{K} \sum_{k=1}^K 
\ln |\Lambda-\lambda_k| +\Lambda x \right \}
-\ln x -1, \cr
&=&\mathop{\rm Extr}_{\Lambda}\left \{
-\int d \lambda \rho_{\bsH}(\lambda)
\ln |\Lambda-\lambda| +\Lambda x \right \}
-\ln x -1. 
\label{G_H_expression1}
\end{eqnarray}
In conjunction with equation (\ref{G_function}), this indicates that 
$G(x)$ is offered as
\begin{eqnarray}
G(x)=\mathop{\rm Extr}_{\Lambda}\left \{
-\int d \lambda \rho(\lambda)
\ln |\Lambda-\lambda| +\Lambda x \right \}
-\ln x -1, 
\label{G_expression1}
\end{eqnarray}
using average spectrum $\rho (\lambda)=\left [ \rho_{\bsH} 
(\lambda) \right ]_{\bsH}$
if a self-averaging property $\rho_{\bsH}(\lambda ) \to \rho(\lambda)$ 
as $K,L \to \infty$ $(\beta =K/L\sim O(1))$ 
holds for typical samples of $\bH$. 

$\rho(\lambda)$ can be formally assessed as follows \cite{Opper1989}. 
For this, we introduce a partition function of 
complex Gaussian spins as 
\begin{eqnarray}
Z^{\rm Gauss}_{\bsH}(\lambda)&\equiv& \int_{\mC^{K}} d\bu \exp \left [-\bu^\dagger
(\Lambda \bI_K-\bH^\dagger\bH )\bu \right ] \cr
&=& \pi^K \left [\det(\Lambda \bI_K-\bH^\dagger\bH ) \right ]^{-1}, 
\label{gauss_partition}
\end{eqnarray}
where $\bu$ is a $K$-dimensional complex vector and 
$\bI_K$ denotes a $K \times K$ identity matrix. 
The dispersion formula 
\begin{eqnarray}
\delta(\Lambda-\lambda)=\lim_{\epsilon \to +0}
\frac{1}{\pi} {\rm Im}\left (\frac{1}{\Lambda-\lambda+i \epsilon}\right )
=-\frac{1}{\pi} {\rm Im}\left (
\frac{\partial }{\partial \Lambda}
\ln (\Lambda-\lambda) \right ), 
\label{dispersion}
\end{eqnarray}
indicates that $\rho(\lambda)$ can be assessed as 
\begin{eqnarray}
\rho(\Lambda)=\frac{1}{\pi } {\rm Im} \left [ 
\frac{\partial}{\partial \Lambda}
\frac{1}{K}\left [ \ln Z^{\rm Gauss}_{\bsH}(\Lambda) \right ]_{\bsH}
\right ], 
\label{replica_rho}
\end{eqnarray}
where $K^{-1}\left [ \ln Z^{\rm Gauss}_{\bsH}(\Lambda) \right ]_{\bsH}$ 
can be evaluated by the replica method. 

For this evaluation, we assess the moments of 
$Z^{\rm Gauss}_{\bsH}(\lambda)$ for $n \in \mN$ 
with use of the saddle-point method, assuming that 
the saddle point is characterized by the Hermitian matrix
$\bQ=(Q_{ab})=(K^{-1}\bu^a\cdot \bu^b)$
$(a,b=1,2,\ldots,n)$. 
This yields the following expression: 
\begin{eqnarray}
&&\frac{1}{K}\ln 
\left [ \left (Z^{\rm Gauss}_{\bsH}(\Lambda) \right )^n \right ]_{\bsH} \cr
&&=\mathop{\rm Extr}_{\hat{\bsQ}, \bsQ}
\left \{ 
\frac{1}{K} 
\left [
\ln 
\int_{\mC^{nK}} \prod_{a=1}^n 
d \bu^a \exp \left [
-\sum_{a=1}^n 
(\bu^a)^\dagger \left ((\Lambda-\hat{Q}_{aa}) \bI_K-\bH^\dagger \bH \right )
\bu^a \right . \right . \right .\cr
&& \left . \left . \left . 
\phantom{aaaaaaaaaaaaaaaaaaaaaaaaaaa} +\sum_{a \ne b} \hat{Q}_{ab} \bu^a 
\cdot \bu^b \right ] \right ]_{\bsH}  -{\rm Tr} \hat{\bQ}\bQ 
\right \}, 
\label{Gauss_replica}
\end{eqnarray}
where $\hat{\bQ}=(\hat{Q}_{ab})$ denotes the conjugate matrix of 
$\bQ$ (Hermitian: does not indicate the Hermitian conjugate of $\bQ$) to 
perform the saddle-point assessment. A distinctive 
property of equation (\ref{Gauss_replica}) is that 
$\hat{\bQ}=0$ always satisfies the saddle-point condition 
for $\forall{n} \in \mN$. 
This means that 
\begin{eqnarray}
&&\frac{1}{nK}\ln 
\left [ \left (Z^{\rm Gauss}_{\bsH}(\Lambda) \right )^n \right ]_{\bsH} 
=\frac{1}{K} 
\left [
-\ln \det \left (\Lambda \bI_K-\bH^\dagger \bH 
\right )
\right ]_{\bsH} + \ln \pi \cr
&&=\frac{1}{K} 
\ln \left [Z^{\rm Gauss}_{\bsH}(\Lambda) \right ]_{\bsH}, 
\label{quench_anneal}
\end{eqnarray}
generally holds for the partition function
and $\forall{n} \in \mN$. Extending this 
expression analytically from $n \in \mN$ to $n \in {\bf \mR}$ and taking $n \to 0$
yield a formula by which to evaluate $\rho (\lambda)$
using the annealed average of the partition function as
\begin{eqnarray}
\rho(\Lambda)=\frac{1}{\pi } {\rm Im} \left [ 
\frac{\partial}{\partial \Lambda}
\frac{1}{K}\ln \left [ Z^{\rm Gauss}_{\bsH}(\Lambda) \right ]_{\bsH}
\right ]. 
\label{anneal_rho}
\end{eqnarray}
Let us insert this expression to equation 
(\ref{G_expression1}) and perform partial integral. 
This operation and the dispersion formula (\ref{dispersion}) yield another 
expression of $G(x)$ as follows:
\begin{eqnarray}
&&G(x)=\mathop{\rm Extr}_{\Lambda}
\left \{ \frac{1}{K} \ln \left [Z_{\bsH}^{\rm Gauss}(\Lambda) \right ]_{\bsH}
+\Lambda x -\ln \pi \right \}-\ln x-1 \cr
&&=\mathop{\rm Extr}_{\Lambda}
\left \{
\frac{1}{K} \ln \int_{\mC^{K}} \frac{d\bu}{\pi^K}
\left [ \exp \left [
-\bu^\dagger (\Lambda \bI_K -\bH^\dagger \bH) \bu \right ]
\right ]_{\bsH} + \Lambda x \right \}-\ln x-1 \cr
&&=\frac{1}{K}
\ln \left [ \frac{\int_{\mC^{K}} d \bu \delta(|\bu|^2-Kx) \exp \left [ \bu^\dagger
(\bH^\dagger\bH) \bu \right ]}
{\int_{\mC^{K}} d \bu \delta(|\bu|^2-Kx)}
\right ]_{\bsH} \cr
&&= \frac{1}{K} \ln \left [ \exp \left [KG_{\bsH}(x) \right ]
\right ]_{\bsH}. 
\label{G_expression2}
\end{eqnarray}
Equations (\ref{G_function}), (\ref{G_H_expression1}), and 
(\ref{G_expression2}) indicate that 
equivalence between annealed and quenched averages
holds in the assessment of $G(x)$, which validates replacement in 
equation (\ref{anneal_quench_G}).

The current argument may be useful in assessing the typical performance of an ensemble of channels. Equations (\ref{cauchy}) and (\ref{R_transform}) can be used for evaluating the performance of a single sample of $\bH$ or for evaluating the performance of a channel ensemble, in which the eigenvalue spectrum is fixed \cite{TakedaUdaKabashima2006,MullerGuoMoustakas2007}. However, naive extension to the analysis of the typical performance of a channel ensemble along this direction generally requires taking configurational averages with respect to a certain distribution of $\bH$ {\em after } evaluating the sample-by-sample performance of $\bH$ using equations (\ref{cauchy}) and (\ref{R_transform}). From a practical viewpoint, this is not possible. However, equation (\ref{G_expression2}) indicates that the typical performance of a channel ensemble can be evaluated using a single function $G(x)$ that characterizes the ensemble property if the eigenvalue spectra are self-averaging. This function can be assessed by the calculation of an annealed average, which, practically speaking, is in most cases much simpler than the calculation of the quenched averages. Formula (\ref{anneal_rho}) is useful as well because this equation can be used to numerically evaluate $\rho(\lambda)$ for a certain class of ensemble, for which analytical evaluation of $\rho(\lambda)$ is difficult. This is demonstrated in the next section. 

%%%%%%%%%%%%%%%%%%%%%%%%%%%%%%%%%
\section{Application}
\subsection{Kronecker model}
Let us demonstrate the significance of the proposed framework 
by analyzing a certain channel ensemble. The ensemble that 
we focus on is termed the Kronecker model, which is a standard MIMO model \cite{TulinoVerdu2004, ShiuFoshiniGansKahn2000, ChizhikFarrokhiLingLozano2000}. 

In this model, the channel transfer matrix $\bH$ is 
represented as 
\begin{eqnarray}
\bH=\sqrt{\bR_r} \bZ \sqrt{\bR_t}, 
\label{Kronecker}
\end{eqnarray}
where $\bZ=(z_{lk})$ is an $L \times K$ random matrix,
each component of which is independently 
sampled from an identical circularly symmetric complex
Gaussian distribution $P(z) =L \pi^{-1} \exp \left [-L |z|^2 \right ]$. 
Here, $\bR_r$ is an $L \times L$ Hermitian matrix that 
represents the effect of spatial correlation among receivers
and antennas. The $K \times K$ Hermitian matrix $\bR_t$ 
represents similar effects for transmit antennas. 
We assume that eigenvalue spectra of $\bR_r$ and $\bR_t$ are
given as $\rho_r(\lambda)$ and $\rho_t(\lambda)$, respectively. 

\subsection{Average eigenvalue spectrum }
For analyzing the typical property, we first introduce 
an expression of partition function of the complex Gaussian spins 
\begin{eqnarray}
&& Z_{\bsH}^{\rm Gauss}(\Lambda)=
\int_{\mC^{K}} d\bu \exp \left [-\bu^\dagger
(\Lambda \bI_{K} -\bH^\dagger\bH) \bu \right ] \cr
&&=\int_{\mC^{K}} d\bu \exp \left [-\Lambda|\bu|^2
+(\bZ \sqrt{\bR_t} \bu)^\dagger \bR_r (\bZ \sqrt{\bR_t} \bu) 
\right ]. 
\label{Kronecker_Gauss}
\end{eqnarray}
Note that 
for fixed $\bu$ in the integrand of equation (\ref{Kronecker_Gauss})
$\bZ \sqrt{\bR_t} \bu\equiv \bv=(v_l)
(l=1,2,\ldots,L)$
can be handled as a circularly symmetric zero-mean $L$-dimensional 
complex Gaussian random vector that is characterized
by covariance $\left [ v_l^*v_j \right ]_{\bsH}
=L^{-1} \bu^\dagger \bR_t \bu \delta_{lj}
=K^{-1} \beta \bu^\dagger \bR_t \bu \delta_{lj}$
$(l,j =1,2,\ldots,L)$
when each component of $\bZ$ in equation (\ref{Kronecker})
is independently sampled from the identical 
distribution $P(z) =L \pi^{-1} \exp \left [-L |z|^2 \right ]$. 
Taking this property into consideration, we evaluate
the annealed average of equation (\ref{Kronecker_Gauss}) as
\begin{eqnarray}
&&\left [Z_{\bsH}^{\rm Gauss}(\Lambda)\right ]_{\bsH} 
=\int d(KQ) d (K\beta T)
\int_{\mC^{K}} d\bu \delta(|\bu|^2-KQ)
\delta(\bu^\dagger \bR_t \bu-K\beta T) \cr
&&\phantom{aaaaaaaaa}\times \exp \left [-\Lambda|\bu|^2 \right ]\times 
\int_{\mC^{L}} \frac{d\bv}{(\pi \beta T)^N}
\exp \left [-\frac{|\bv|^2}{\beta T}+\bv^\dagger \bR_r \bv \right ] \cr
&&= \int d(KQ) d (K\beta T)
\int_{-i \infty}^{i \infty} 
\frac{d\hat{Q}}{2\pi i}
\frac{d\hat{T}}{2\pi i}
\int_{\mC^{K}} d\bu 
\exp \left [-\bu^\dagger ((\Lambda+\hat{Q})\bI_K-\hat{T}\bR_t )
\bu \right ] \cr
&&\phantom{aaaaaaaaa}\times \exp \left [K(\hat{Q}Q-\hat{T}T) \right ]
\times \left [\det (\bI_L-\beta T \bR_r ) \right ]^{-1} \cr
&&=\int d(KQ) d (K\beta T)
\int_{-i \infty}^{i \infty} 
\frac{d\hat{Q}}{2\pi i}
\frac{d\hat{T}}{2\pi i}
\exp \left [K(\hat{Q}Q-\hat{T}T) \right ] \cr
&&\phantom{aaaaaaaaa}\times 
\pi^K  \left [ \det ((\Lambda+\hat{Q}) \bI_{L}-\hat{T}\bR_t )
\right ]^{-1}
\times \left [\det (\bI_L-\beta T \bR_r ) \right ]^{-1}. 
\end{eqnarray}
Utilizing the relations 
$K^{-1} \ln \det ((\Lambda+\hat{Q}) \bI_{K}-\hat{T}\bR_t )
=\int d \lambda \rho_t(\lambda)\ln (\Lambda+\hat{Q}-\hat{T}\lambda)$
and $L^{-1} \ln \det (\bI_{L}-\beta T\bR_r )
=\int d \lambda \rho_r(\lambda)\ln (1-\beta T \lambda)$, 
this equation indicates that the annealed average of the partition 
function is evaluated by the saddle-point method as
\begin{eqnarray}
&&\frac{1}{K}\ln \left [Z_{\bsH}^{\rm Gauss}(\Lambda)\right ]_{\bsH} \cr
&&= \mathop{\rm Extr}_{\hat{Q},Q,\hat{T},T}
\left \{ \ln \pi +\hat{Q}Q-\hat{T}T \right . \cr
&& \left .
\phantom{aaaaaa}
-\int d \lambda \rho_t(\lambda)\ln (\Lambda+\hat{Q}-\hat{T}\lambda)
- \frac{1}{\beta}
\int d \lambda \rho_r(\lambda)\ln (1-\beta T \lambda) 
\right \}. 
\label{saddle_point_Kronecker}
\end{eqnarray}
This yields the following equations: 
\begin{eqnarray}
T&=&\int d\lambda \frac{\lambda \rho_t(\lambda)}{\Lambda-\hat{T}\lambda}, 
\label{Kronecker_SP1}\\
\hat{T}&=&\int d\lambda \frac{\lambda \rho_r(\lambda)}{1-\beta T \lambda}, 
\label{Kronecker_SP2}
\end{eqnarray}
which are only relevant for the saddle-point condition because 
$\hat{Q}=0$ always holds. 
The solution of equations (\ref{Kronecker_SP1}) and (\ref{Kronecker_SP2}) yields
the average eigenvalue spectrum as
\begin{eqnarray}
\rho(\Lambda)=\frac{1}{\pi}{\rm Im}(Q)
=\frac{1}{\pi} {\rm Im} \left \{
\int d \lambda \frac{\rho_t(\lambda)}{\Lambda-\hat{T}\lambda}
\right \}. 
\label{Kronecker_rho}
\end{eqnarray}

\subsection{Performance assessment for correlation 
matrices of T\"{o}plitz type}

We applied the proposed scheme to the case in which 
both correlation matrices $\bR_r$ and $\bR_t$ are of the T\"{o}plitz type, 
in which the $(i,j)$ elements of $\bR_r$ and $\bR_t$ are 
given as $r^{|i-j|}$ and $t^{|i-j|}$, respectively, 
where $0<r<1$ and $0<t<1$. For matrices of this type, 
eigenvalue spectra can be computed analytically in the 
limit of $K,L \to \infty$ as 
\begin{eqnarray}
\rho_r(\lambda)=\frac{1}{\pi \lambda \sqrt{(\alpha_{+}-\lambda)
(\lambda-\alpha_-)}}, 
\quad \alpha_{\pm }=\frac{1\pm r}{1\mp r}, 
\label{Toeplitz_dist}
\end{eqnarray}
and similarly for $\rho_t(\lambda)$. 
For the purpose of illustrating generality, we employed numerical methods 
based on equations (\ref{G_expression1}) and (\ref{anneal_rho})
to analyze this model, although expression (\ref{Toeplitz_dist}) 
potentiates further analytical treatment, 
which will be reported elsewhere \cite{HatabuTakedaKabashima2007}. 

\begin{figure}[t]
\setlength{\unitlength}{1mm}
\begin{picture}(1870,69)
\put(41,64){\includegraphics[width=60mm,angle=270]{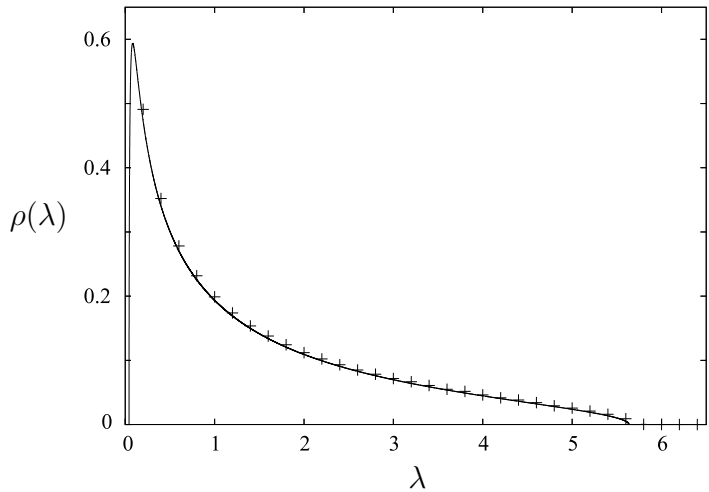}}
\put(32,35){$\rho(\lambda)$}
\put(85,0){$\lambda$}
\end{picture}
\caption{
 Eigenvalue spectrum of channel transfer cross-correlation
 matrix $\bH^{\dagger} \bH$.
 The comparison of the analytical result obtained 
 by the scheme demonstrated in Section 4 (curve) and the result obtained 
 by numerical diagonalization of randomly-generated cross-correlation
 matrix (markers) is depicted.  Here, each parameter is set as $r=t=0.2, \beta=1.5$,
 and 200 samples of a $500 \times 750$ random channel transfer matrix are 
 used for numerical diagonalization. The results show good agreement.}
\label{fig1}
\end{figure}
\begin{figure}[t]
\setlength{\unitlength}{1mm}
\begin{picture}(180,67)
\put(8,56){\includegraphics[width=50mm,angle=270]{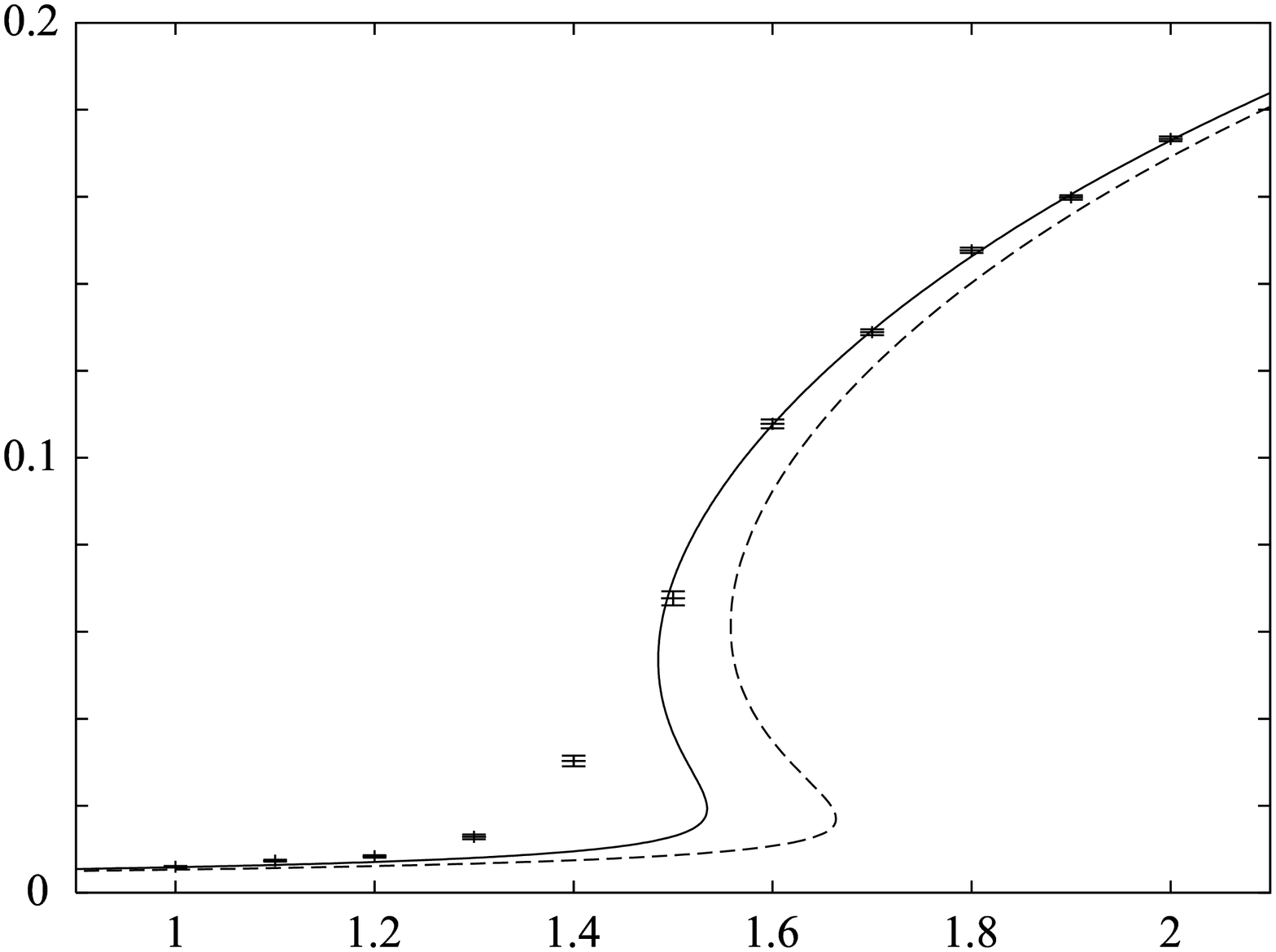}}
\put(2,32){$P_b$}
\put(34,1){$\beta=K/L$}
\put(1,56){(a)}
\put(88,56){\includegraphics[width=50mm,angle=270]{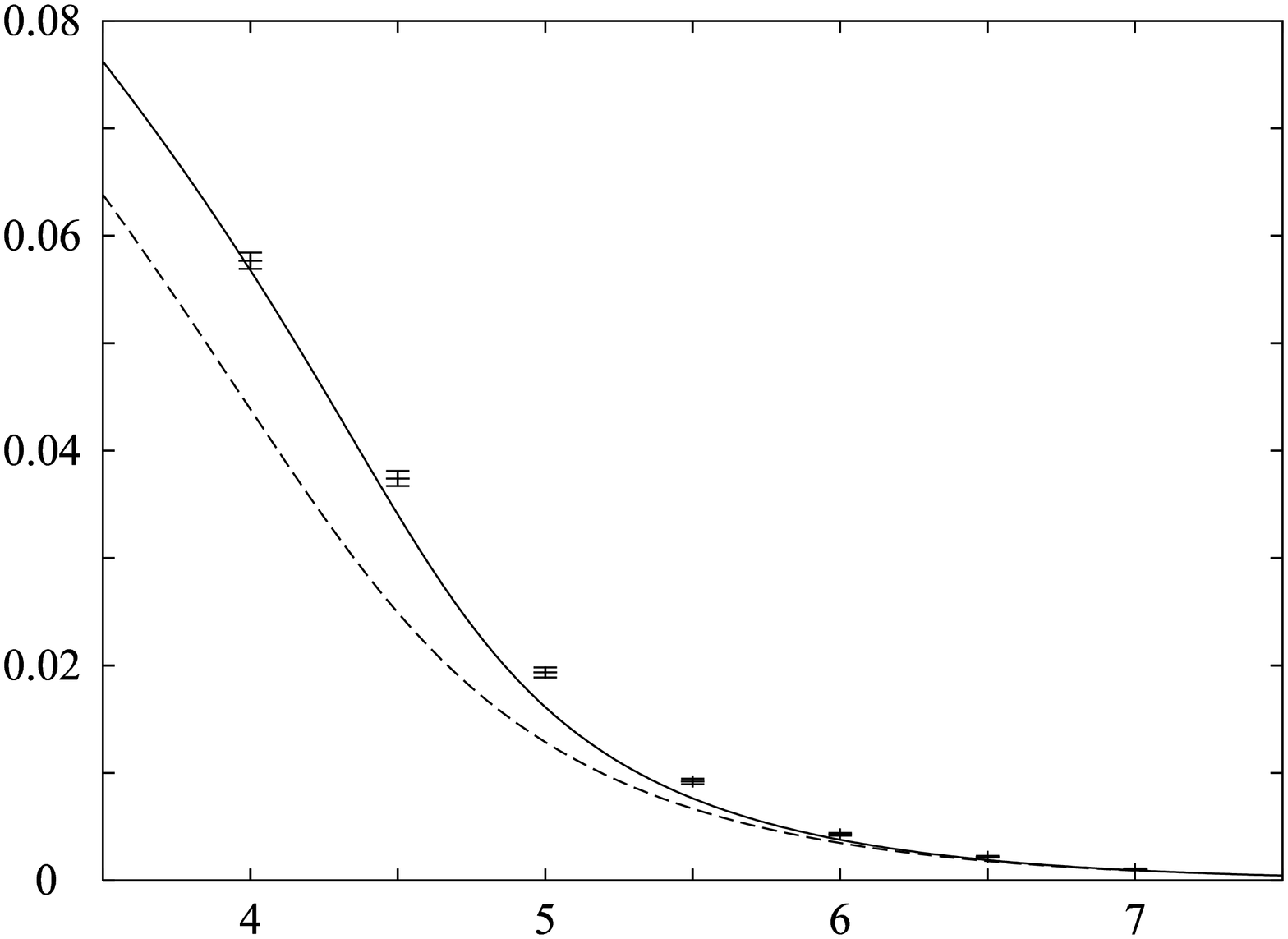}}
\put(81,32){$P_b$}
\put(115,1){$E_b/N_0$ [dB]}
\put(80,56){(b)}
\end{picture}
\caption{
(a): Bit error rate $P_b$ vs. the ratio $\beta=K/L$
under the condition $N_0=N_r=0.37, L=1024$, and a $500$ sample average. 
The solid/broken lines indicate the theoretical prediction
obtained by the proposed scheme for the cases of $r=t=0.2$/$r=t=0$
 (uncorrelated), respectively. The bars indicate the result 
 of the numerical experiment in the case of $r=t=0.2$,
 which shows good agreement with the prediction, except for the region in the vicinity of the waterfall (first-order phase transition) point.
(b): Bit error rate $P_b$ vs. signal-to-noise ratio $E_b/N_0$ under
the condition of $\beta=1.1, r=t=0.2, L=1024$, and a $500$ sample average. The bars, which express the numerical results, show good agreement with the theoretical
 prediction (solid). The broken line indicates the analytical results
 for the uncorrelated case.}
\label{fig2}
\end{figure}

Figure \ref{fig1} shows a comparison between the numerical average of the eigenvalue spectrum estimated from 200 samples of a $500 \times 750$ random channel transfer matrix (markers) and the theoretical prediction calculated from equation (\ref{Kronecker_rho}) by numerically solving equations (\ref{Kronecker_SP1}) and (\ref{Kronecker_SP2}) (curve). The excellent agreement between the experimental and theoretical data in this figure indicates that our approach to the assessment of the average eigenvalue spectrum based on the numerical saddle-point analysis of the annealed average of a partition function works rather efficiently in this case. 

Figures \ref{fig2}(a) and \ref{fig2}(b) represent the communication performance assessed by the current framework. For convenience, the data in these figures are computed for $S=2$ (BPSK) and {\em real} channels in which all elements of $\bH$ and channel noises are restricted to real numbers by replacing complex normal distribution $P(u)=\pi^{-1} \exp \left [-|u|^2 \right ]$ and unitary matrix $\bU$ with real normal distribution $P(x)=(2\pi)^{-1/2} \exp \left [-x^2/2 \right ]$ and orthogonal matrix $\bO$ in computing $G(x)$. Even if such a restriction is imposed, the framework is completely parallel. The only difference is that prefactor $1/2$ is placed in front of the definitions of $G_{\bsH}(x)$ and $G(x)$. The curves are provided by the saddle-point condition of equation (\ref{free_energy}), numerically assessing $G(x)$ from $\rho(\lambda)$ using equation (\ref{G_expression1}), while the experimental results denoted by markers were obtained by the Thouless-Anderson-Palmer equation of the proposed system based on a strategy similar to equation (12) of ref. \cite{TakedaUdaKabashima2006}. The reasonably good consistency between the curves and markers indicates the significance of the proposed framework in the performance assessment of linear vector channels of discrete inputs. 

%%%%%%%%%%%%%%%%%%%%%%%%%%%%%%%%%
\section{Summary}
In summary, we have developed a framework by which to analyze linear vector channels of discrete inputs based on the replica method. 
Assuming uniformity of the prior probability of inputs
makes it possible to characterize the typical property 
of the objective channel ensemble by a single 
function $G(x)$ if the eigenvalue spectrum of 
the cross-correlation matrix of the channel transfer matrix
is self-averaging. 
We have also presented a practical scheme by which to evaluate 
$G(x)$ and an average eigenvalue spectrum $\rho(\lambda)$
using an annealed average of a partition 
function of Gaussian spins. The significance of the proposed scheme is demonstrated by application to an existing channel ensemble called the Kronecker model. 

Future studies will include the mathematical validation of equation (\ref{replacement}) and the development of a practical demodulation algorithm \cite{Kabashima2003,TanakaOkada2005}.

\ack

This study was supported in part by Grants-in-Aid from JSPS/MEXT, Japan
(Nos. 17340116 and 18079006).

\appendix
\section{Free energy under 1RSB ansatz}
Under the one-step replica symmetry (1RSB) ansatz, 
$n$ replicated replicas $\btau^1,\btau^2, \ldots,\btau^n$ are divided into 
$n/x$ groups of identical size $x$, and 
the relevant saddle point is characterized by relative configuration 
\begin{eqnarray}
\btau^a \cdot \btau^b =
\left \{
\begin{array}{ll}
K, & a=b, \cr
K(q+\Delta), & \mbox{$a$ and $b$ belong to an identical group}, \cr
Kq, & \mbox{otherwise},
\end{array}
\right .
\end{eqnarray}
and $\bone \cdot \btau^a=m$ $(a,b=1,2,\ldots,n)$, 
where $q$, $\Delta$, and $m$ are assumed to be real to satisfy 
obvious symmetry $\btau^a \cdot \btau^b =\btau^b \cdot \btau^a$. 
Here, $x$ serves as Parisi's replica symmetry breaking parameter 
after analytical continuation. 
For $\{\btau^a\}$ that satisfies this configuration, 
$\bL(n)$ has four types of eigenvalues:
$\lambda_1=-K(N_r+nN_0)^{-1}(1-q+(x-1)\Delta+n(1-2m+q))$, 
$\lambda_2=-KN_r^{-1}(1-q+(x-1)\Delta)$, 
$\lambda_3=-KN_r^{-1}(1-q-\Delta)$ and $\lambda_4=0$, the 
degeneracies of which are
$1$, $n/x-1$, $n-n/x$, and $K-n$, respectively. This provides
the following expression: 
\begin{eqnarray}
&&{\rm Tr}G\left (\frac{\bL(n)}{K} \right ) \cr
&&\!\!\!\!\!\!\!\!\!\!
=nG\left (-\frac{1-q-\Delta}{N_r} \right ) 
+\frac{n}{x}\left (
G\left (-\frac{1-q+(x-1)\Delta}{N_r} \right )-
G\left (-\frac{1-q-\Delta}{N_r} \right ) 
\right ) \cr
&&\!\!\!\!\!\!\!\!\!\!
+G\left (-\frac{1-q+(x-1)\Delta+n(1-2m+q)}{N_r+nN_0} \right ) 
-G\left (-\frac{1-q+(x-1)\Delta}{N_r} \right ). \cr
&&
\label{G_1RSB}
\end{eqnarray}
Using the saddle-point method, the number of microscopic configurations
that satisfy the current ansatz increases as $\exp \left [K S_n^{\rm 1RSB}
(q,\Delta,m;x) \right ]$, where
\begin{eqnarray}
&&S_n^{\rm 1RSB}(q,\Delta,m;x) \cr
&&\!\!\!\!\!\!\!\!\!\!
=\mathop{\rm Extr}_{\hat{q},\hat{\Delta},\hat{m}} \left \{
\ln \left[ \int_{\mC}D\zeta 
\left (
\int_{\mC} D\eta \left (
\mathop{\rm Tr}_{\tau \in A_{\cal S}} 
\exp \left [ {\rm Re}\left ((\sqrt{\hat{\Delta}}\eta^*
+\sqrt{\hat{q}}\zeta^* + \hat{m} )
\tau \right ) \right ]  
\right )^x \right )^{\frac{n}{x}} \right ] \right . \cr
&&  -n \hat{m}m -\frac{n}{2}(\hat{q}+\hat{\Delta}) \cr
&& \left . - \frac{n}{x} \frac{x(x-1)}{2}
\left ((\hat{q}+\hat{\Delta})(q+\Delta)-\hat{q}q \right )
-\frac{n(n-1)}{2} \hat{q}q \right \}. 
\label{S_1RSB}
\end{eqnarray}
Equations (\ref{G_1RSB}) and (\ref{S_1RSB}) indicates that under the 1RSB ansatz, 
free energy is assessed as 
\begin{eqnarray}
&&\frac{1}{K} \overline{\ln Z}= \lim_{n \to 0}\frac{1}{nK}\ln
 \overline{Z^n} \cr
&&=\mathop{\rm Extr}_{q,\Delta,m,\hat{q},\hat{\Delta},\hat{m}}
\left \{
G\left (-\frac{1-q-\Delta}{N_r} \right ) \right . \cr
&&+ 
\frac{1}{x}
\left (G\left (-\frac{1-q+(x-1)\Delta}{N_r} \right )
-G\left (-\frac{1-q-\Delta}{N_r} \right ) \right )  \cr
&& + \left (-\frac{1-2m+q}{N_r}+\frac{N_0(1-q+(x-1)\Delta)}{N_r^2} \right )
G^\prime \left (-\frac{1-q+(x-1)\Delta}{N_r} \right ) \cr
&&
-\hat{m}m-\frac{1}{2} (\hat{q}+\hat{\Delta})
- \frac{x-1}{2}
\left ((\hat{q}+\hat{\Delta})(q+\Delta)-\hat{q}q \right )
+\frac{1}{2} \hat{q}q  \cr
&&
\left . +\frac{1}{x}
\int_{\mC}D\zeta \ln \left[ 
\int_{\mC} D\eta \left (
\mathop{\rm Tr}_{\tau \in A_{\cal S}} 
\exp \left [ {\rm Re}\left ((\sqrt{\hat{\Delta}}\eta^*
+\sqrt{\hat{q}}\zeta^* + \hat{m} )
\tau \right ) \right ]  
\right )^x  \right ] \right \}. \cr
&& 
\end{eqnarray}

\end{document}